\documentclass[reprint,amsmath,amssymb,nofootinbib,aps,prc]{revtex4-1}
\usepackage{bm}
\usepackage{caption}
\usepackage{color}
\usepackage{epstopdf}
\usepackage{graphicx}
\usepackage{lmodern}
\usepackage{mathtools}
\usepackage{multirow}
\usepackage{makecell}
\usepackage{subfigure}
\usepackage{textcomp}
\usepackage{ulem}
\usepackage{url}

\begin{document}

\title{Charge-dependent transverse momentum and its impact on the search for the chiral magnetic wave}

\author{Wen-Ya Wu$^{a}$}
\author{Chun-Zheng Wang$^{b,c}$}
\author{Qi-Ye Shou$^{a}$} \thanks{shouqiye@fudan.edu.cn}
\author{Yu-Gang Ma$^{a,b}$} \thanks{mayugang@fudan.edu.cn}
\author{Liang Zheng$^{d}$}
\affiliation{$^a$Key Laboratory of Nuclear Physics and Ion-beam Application (MOE), Institute of Modern Physics, Fudan University, Shanghai 200433, China}
\affiliation{$^b$Shanghai Institute of Applied Physics, Chinese Academy of Sciences, Shanghai 201800, China}
\affiliation{$^c$University of Chinese Academy of Sciences, Beijing 100049, China}
\affiliation{$^d$School of Mathematics and Physics, China University of Geosciences (Wuhan), Wuhan 430074, China}

\begin{abstract}
The chiral magnetic wave (CMW) is sought using the charge asymmetry ($A_{\rm ch}$) dependence of anisotropic flow in heavy-ion collisions. The charge dependent transverse momentum ($p_{\rm T}$), however, could play a role as a background. With the string fragmentation models, including PYTHIA, we demonstrate the origin of the $A_{\rm ch}-p_{\rm T}$ correlation and its connection with the local charge conservation (LCC). The impact of $A_{\rm ch}-p_{\rm T}$ and its behavior in varied kinematic windows are also discussed. This study provides more insights for the search for the CMW and comprehending the collective motion of the quark-gluon plasma.
\end{abstract}

\maketitle

%%%%%%%%%%%%%%%%%%%%%%%%%%%%%%%%%%%%%%%%%%%%%%%%%%%%%%%%%
\section{Introduction} \label{sec:intro}

In relativistic heavy-ion collisions, the interplay of the chiral anomaly and the intense magnetic field created in off-central collisions is proposed to generate several kinds of anomalous chiral phenomena~\cite{Kharzeev:2008, Kharzeev:2016, Hattori:2017, Zhao:2019}, e.g. the chiral magnetic effect (CME), the chiral separation effect (CSE), and the chiral magnetic wave (CMW)~\cite{Burnier:2011, Burnier:2012, Yee:2014, Taghavi:2015}. While the CME could manifest itself in a finite electric dipole moment with respect to the reaction plane~\cite{Wang:2018}, the CMW is expected to generate an electric quadrupole moment in the quark-gluon plasma (QGP), where the ``poles" (out of plane) and the ``equator" (in plane) of the participant region respectively acquire additional positive or negative charges. Taking advantage of the anisotropic emission of particles in the azimuthal direction, it is feasible to measure the CMW using the charge asymmetry ($A_{\rm ch}$) dependence of elliptic flow ($v_2$) between the positively and negatively charged particles, i.e.,
\begin{equation} \label{eq:1}
\Delta v_{2} \equiv v_{2}^{-} - v_{2}^{+} \simeq rA_{\rm ch},
\end{equation}
where $A_{\rm ch} \equiv (N^{+} -N^{-}) / (N^{+} +N^{-})$ with $N$ denoting the number of particles in a given event, and the slope $r$ is used to quantify the strength. Phenomenological simulations~\cite{Ma:2014, Shen:2019} confirm that the charge separation caused by the CMW is bound to bring about this linear dependence.

Over the past decade, the STAR~\cite{STAR:2015, Shou:2019}, ALICE~\cite{ALICE:2016} and CMS~\cite{CMS:2019} collaborations have performed the measurements at various collision energies and systems. A robust relationship between $v_2$ and $A_{\rm ch}$ is observed and the slope extracted at semi central collisions agrees with the theoretical expectation, seemingly bearing out the CMW theory. Nevertheless, the strikingly similar linear relation is also experimentally observed in p-Pb collisions and for triangular flow ($v_3$)~\cite{CMS:2019}. The direction of the magnetic field is irrelevant to the reaction plane in small system collisions~\cite{CMS:2017, Belmont:2017} and the quadrupole configuration is unable to cause the $A_{\rm ch}$-$v_3$ relation. For that reason, one can assert the existence of the non-CMW background.

Understanding the components of the background and how they contribute to the observable are essential to disentangle the CMW signal. Among several sources~\cite{Bzdak:2013, Voloshin:2014, Campbell:2013, Stephanov:2013, Hatta:2016, Hongo:2017, Zhao:2019a}, the most important one is suggested to be the local charge conservation (LCC) entwined with the collective motion of the QGP. Ref.~\cite{Bzdak:2013} introduces the LCC effect into a hydrodynamic model, which can qualitatively generate the linear relation between $v_2$ and $A_{\rm ch}$, albeit with a smaller slope compared to the data. Reference~\cite{Voloshin:2014} demonstrates some basic features of the LCC with a simple blast wave model and proposes a novel observable, three-particle correlator. Both of these two studies mimic the LCC by forcing the charged particles to emit always in pairs (one positively and one negatively charged) at the same spatial point.  On the other hand, without artificially introducing the charge-conserving pair, AMPT simulation fails to reproduce such a linear relation~\cite{Ma:2014, Shen:2019} and reveals that the contribution from the resonance decay can be either negative or positive depending on the mass~\cite{Xu:2020}.

It is noteworthy that a linear dependence between the mean transverse momentum ($\langle p_{\rm T} \rangle$) and $A_{\rm ch}$ has also been reported in the CMS data~\cite{CMS:2019}, i.e.,
\begin{equation} \label{eq:2}
\Delta \langle p_{\rm T} \rangle \equiv \langle p_{\rm T}^{-} \rangle - \langle p_{\rm T}^{+} \rangle \propto A_{\rm ch}.
\end{equation}
The extracted slope of Eq. (\ref{eq:2}) is found to be consistent in Pb-Pb and p-Pb collisions. It is well known that both $v_2$ and $v_3$ linearly depend on $p_{\rm T}$. Thus, the relationship between $A_{\rm ch}$ and $\langle p_{\rm T} \rangle$ can naturally give rise to the dependence between $A_{\rm ch}$ and $v_n$, serving as a background in the search for the CMW. There are reasons to presume that Eq. (\ref{eq:2}) is a consequence of the LCC~\cite{Voloshin:2014}. Few work, however, have quantitatively established the connection between $A_{\rm ch}$ and $p_{\rm T}$ in a realistic environment. In this work, we concentrate on the origin of this charge dependent $p_{\rm T}$ and investigate the feature of the LCC with the string fragmentation models. We further discuss its impact on the search for the CMW.

%%%%%%%%%%%%%%%%%%%%%%%%%%%%%%%%%%%%%%%%%%%%%%%%%%%%%%%%%
\section{Charge dependent transverse momentum and the local charge conservation} \label{sec:sec2}

\begin{figure}
\centering
\includegraphics[width=\linewidth]{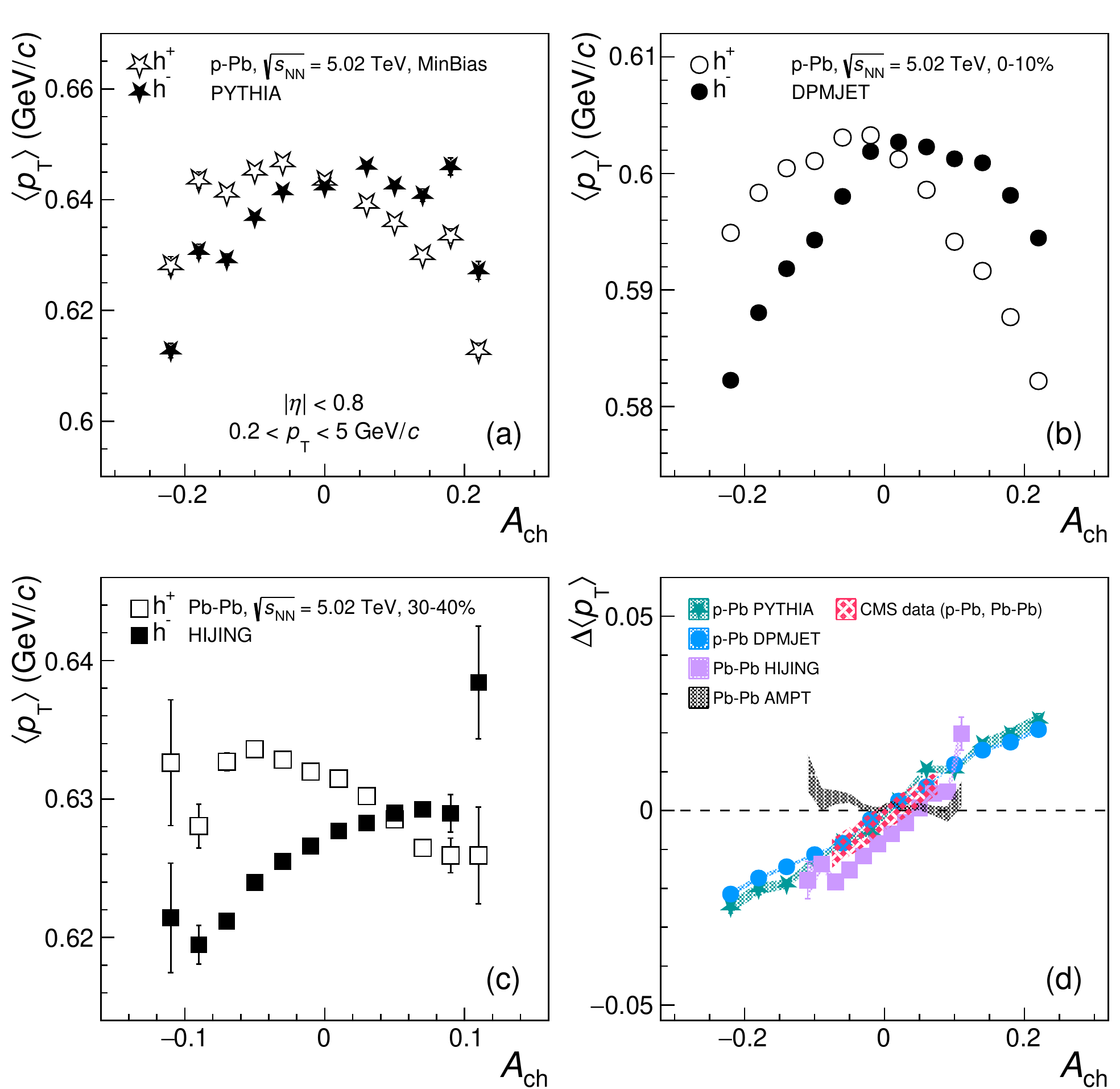}
\captionsetup{justification=raggedright}
\captionof{figure}{Dependence of $\langle p_{\rm T} \rangle$ (a-c) and $\Delta \langle p_{\rm T} \rangle$ (d) on $A_{\rm ch}$ in Pb-Pb and p-Pb collisions at $\sqrt{s_{\rm NN}}$ = 5.02 TeV with different models.}
\label{fig:pt_ach}
\end{figure}

The main models used in this analysis include PYTHIA~\cite{Sjostrand:2015,  Sjostrand:2020}, DPMJET~\cite{Capella:1994} and HIJING~\cite{Gyulassy:1994}. All of them employ the Lund string fragmentation~\cite{Ferreres-Sole:2018} formalism to deal with the hadronization process, although with different machineries at parton level to construct the color singlet string objects. In the string fragmentation picture, the final state hadrons are produced through the iterative breakups of the string system based on the linear confinement assumption. For a simple string object consisting of a quark and an antiquark endpoints, a new quark-antiquark pair can be created during the string breakup in the middle of two endpoints. They must be produced at the same space-time vertex to meet the requirement of local flavor conservation and then pulled apart by the string tension to form two hadrons. Generally, this process begins with low momentum particles in the central region of the string and then spreads outwards to high momentum particles at later times~\cite{Ferreres-Sole:2018}. In addition to the string fragmentation models, the string melting version of the AMPT model~\cite{Lin:2005}, whose initial condition and hadronization are handled by HIJING and a naive quark coalescence respectively, is also adopted as a comparison.

\begin{figure}
\centering
\includegraphics[width=\linewidth]{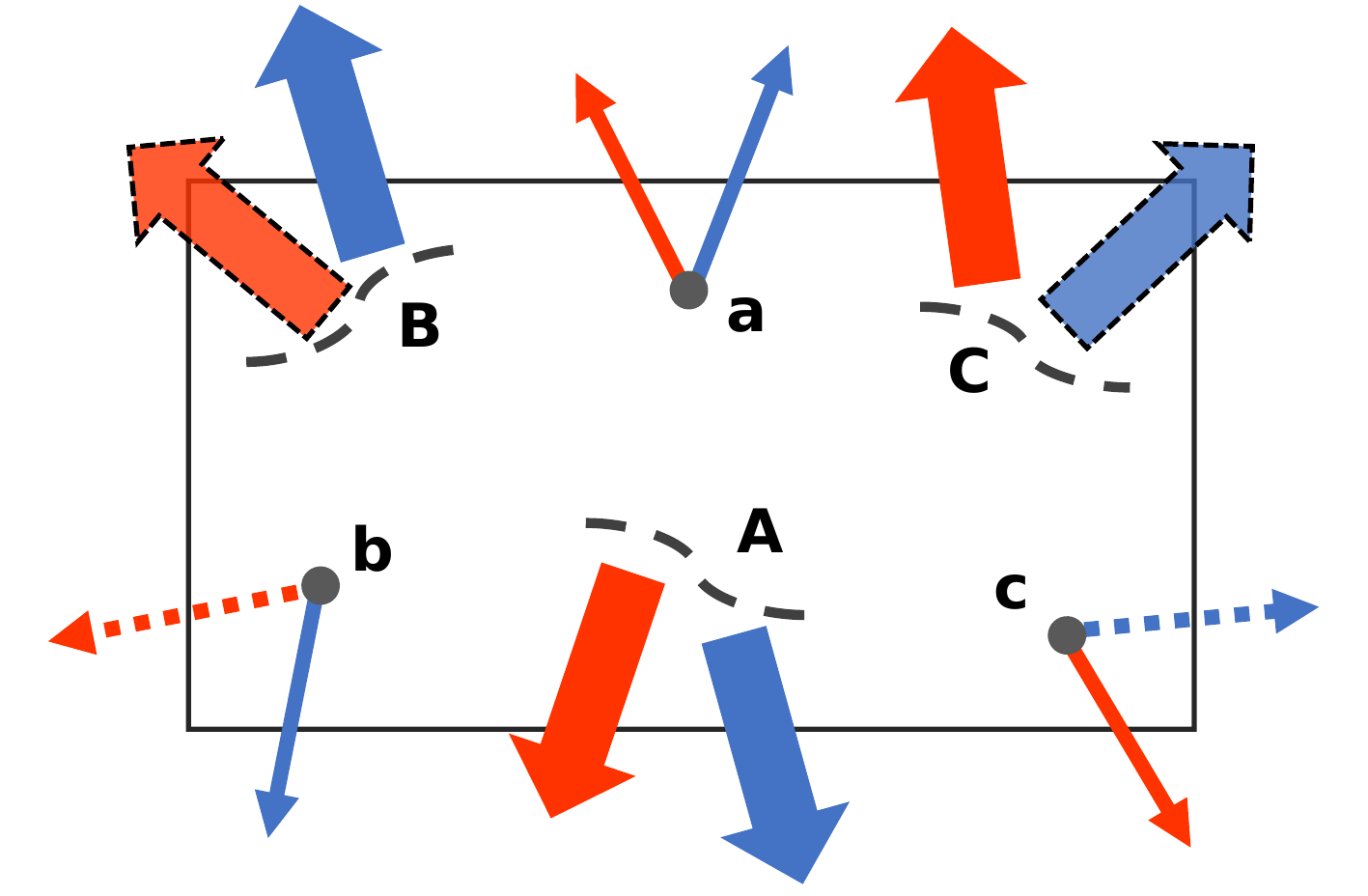}
\captionsetup{justification=raggedright}
\captionof{figure}{A schematic view of a PYTHIA event consisting of six production cases. The rectangle denotes the detector in the longitudinal direction. Capital letters denote the string fragmentation while lowercase letters denote the resonance decay with two daughters. Positive and negative charge are marked in red and blue, respectively.}
\label{fig:sch}
\end{figure}

First we examine the $A_{\rm ch}$ dependence of $\langle p_{\rm T} \rangle$ for positive ($h^+$) and negative ($h^-$) hadrons in p-Pb and Pb-Pb collisions at $\sqrt{s_{\rm NN}}$ = 5.02 TeV with the aforementioned models. $A_{\rm ch}$ is calculated with all charged hadrons at final state with $p_{\rm T} > 0.2$ GeV/$c$ and $|\eta| < 0.8$. A wide $p_{\rm T}$ coverage of $0.2 < p_{\rm T} < 5$ GeV/$c$ is applied to estimate $\langle p_{\rm T} \rangle$, matching the experimental selection criteria. As presented in Fig.~\ref{fig:pt_ach} (a)-(c), the $\langle p_{\rm T} \rangle$ of $h^+$ is systematically larger than that of $h^-$ when $A_{\rm ch} < 0$ and such a trend reverses when $A_{\rm ch} > 0$. This feature is qualitatively consistent with the experimental measurement though the relations between $A_{\rm ch}$ and $\langle p_{\rm T} \rangle$ in the models are not always monotonic. The $\langle p_{\rm T} \rangle$ difference between $h^-$ and $h^+$ is calculated and normalized by their average:
\begin{equation}  \label{eq:3}
\Delta \langle p_{\rm T} \rangle = \frac{p_{\rm T}^{-} - p_{\rm T}^{+}}{(p_{\rm T}^{-} + p_{\rm T}^{+})/2},
\end{equation}
to make the apples-to-apples comparison between different systems. In Fig.~\ref{fig:pt_ach} (d), similar linear dependences with a common normalized slope of $\sim$0.1 can be clearly seen in the PYTHIA, DPMJET and HIJING, which agree perfectly with the CMS measurement~\footnote{The CMS data~\cite{CMS:2019} in p-Pb and Pb-Pb collisions are merged here since they are almost identical.}. In contrast, no dependence is found in the AMPT model as marked by the dark grey band. The reason that such a charge ($q$) - $p_{\rm T}$ correlation exists in the Lund string family models but not AMPT may be attributed to the fact that, in the AMPT, the spatial charge distribution initially stemming from HIJING is largely distorted during the parton rescattering stage implemented by Zhang’s Parton Cascade (ZPC) model. The $q-p_{\rm T}$ correlation is therefore no longer guaranteed when the grouped quarks are hadronized to final state particles through coalescence, as proposed in Ref.~\cite{Du:2007, Li:2009}.

\begin{figure}
\centering
\includegraphics[width=\linewidth]{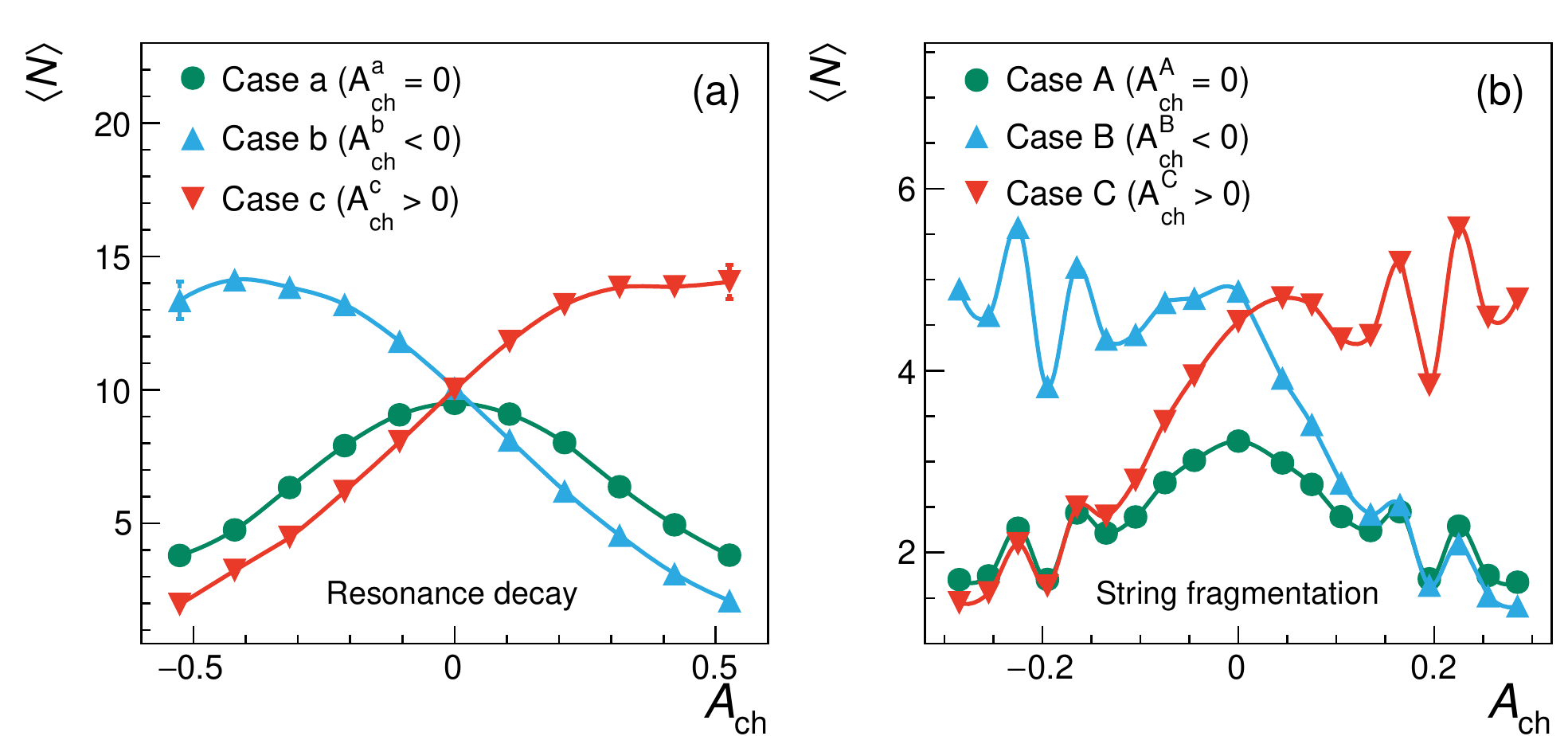}
\captionsetup{justification=raggedright}
\captionof{figure}{The average number of each production case (see Fig.~\ref{fig:sch}) as a function of the event $A_{\rm ch}$ in the PYTHIA p-Pb collisions.}
\label{fig:nCase_ach}
\end{figure}

To understand the intrinsic correlation between $q$ and $p_{\rm T}$, the PYTHIA event is dissected and illustrated in Fig.~\ref{fig:sch}. Regardless of collision systems and energies, final state particles originate in either primordial production (labelled $A$, $B$ and $C$) from the string fragmentation (black dashed curves) or the resonance decay (labelled $a$, $b$ and $c$). For the cases of $A$ and $a$, all particles are emitted within the detector so the total measured charges remain neutral. For the cases of $B$ ($C$) and $b$ ($c$), however, more $h^+$ ($h^-$) escape from the detector due to the limited acceptance, leading to the charge imbalance. A typical event consists of a few (tens) of each cases and the event $A_{\rm ch}$ can be arithmetically decomposed into the weighted sum of the charge asymmetry of each case, $A_{\rm ch}^{i}$:
\begin{equation} \label{eq:4}
\begin{split}
A_{\rm ch} = \frac{1}{M} \sum N^{\it i}m^{\it i}A_{\rm ch}^{\it i},
\end{split}
\end{equation}
where $M$ is the event multiplicity; $N^{\it i}$ and $m^{\it i}$ are the average number and the average multiplicity of the case $i$, respectively, and $\it i$ loop over all six cases. The relation between the event $A_{\rm ch}$ and $N$ is shown in Fig.~\ref{fig:nCase_ach}. It can be seen that the event $A_{\rm ch}$ is mainly determined by the numbers of each production case. Apparently, the more cases of $B$ ($C$) and $b$ ($c$) one event has, the more negative (positeve) the $A_{\rm ch}$ is and vice versa. For those events of $A_{\rm ch} \approx$ 0, the number of $B$($b$) equals to that of $C$($c$), counterbalancing the difference between $A_{\rm ch}^B$ and $A_{\rm ch}^C$, and the proportion of $A$ and $a$ reach the maximum.

\begin{figure}
\centering
\includegraphics[width=\linewidth]{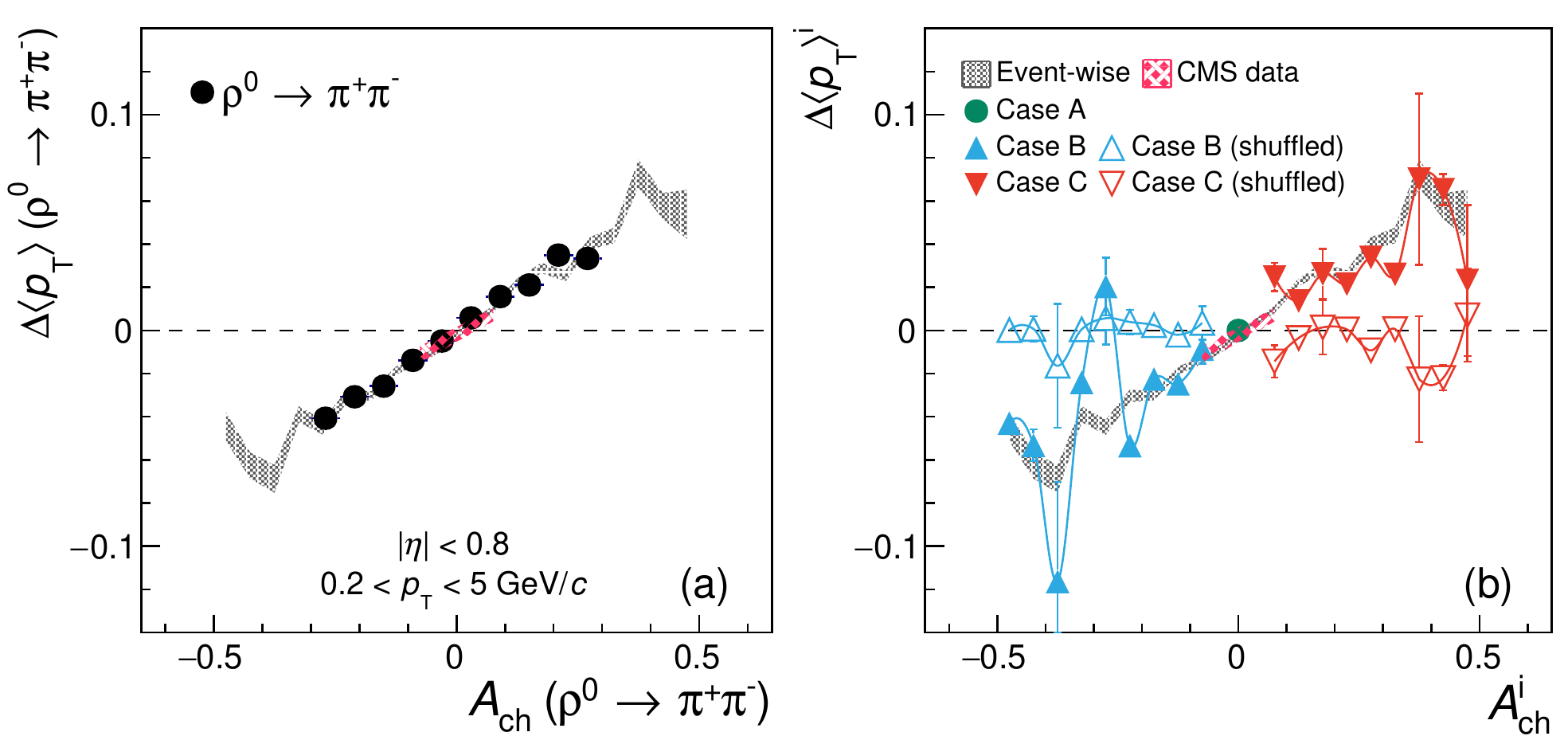}
\captionsetup{justification=raggedright}
\captionof{figure}{The local correlation of $A_{\rm ch}^{\it i}-\Delta \langle p_{\rm T}  \rangle^{\it i}$ from (a) the resonance decay of $\rho^0\rightarrow\pi^+\pi^-$ and (b) the string fragmentation.}
\label{fig:pt_ach_str}
\end{figure}

As per Eq. (\ref{eq:4}), the event-wise $A_{\rm ch}-\Delta \langle p_{\rm T} \rangle$ dependence can be converted into the string and the resonance levels. We start with the contribution from the resonance decay in a given event with $\rho^0\rightarrow\pi^+\pi^-$. Such a typical dynamic process of LCC has proven to be a significant background in the search for the CME~\cite{Zhao:2019, Wang:2018}. In Fig.~\ref{fig:pt_ach_str} (a), we calculate both $A_{\rm ch}$ and $\Delta \langle p_{\rm T} \rangle$ with the decayed $\pi^{+/-}$ and find a linear dependence with a positive slope, which is very consistent with the event-wise $A_{\rm ch}-\Delta \langle p_{\rm T} \rangle$ and the CMS data. This can be easily understood by the following derivation:
\begin{equation} \label{eq:5}
\begin{aligned}
& \Delta \langle p_{\rm T} \rangle
= \frac{p_{\rm T}^{a}m^{a} + p_{\rm T}^{b}m^{b}}{m^{a}+m^{b}} - \frac{p_{\rm T}^{a}m^{a} + p_{\rm T}^{c}m^{c}}{m^{a}+m^{c}} \\
& = \frac{m^{a}\ (m^{c} - m^{b})\ (p_{\rm T}^{a} - p_{\rm T}^{b})}{(m^{a}+m^{b})(m^{a}+m^{c})},
\end{aligned}
\end{equation}
where $m^i$ is the multiplicity of case $i$. Obviously decayed $\pi$ emitted near the edge of the detector are more likely to be unpaired and carry smaller $p_{\rm T}$, i.e., $p_{\rm T}^{\rm a}$ (0.64) $> p_{\rm T}^{\rm b} \approx p_{\rm T}^{\rm c}$ (0.59), as shown in Table~\ref{tab:tab1}. Thus, for those events of $A_{\rm ch}^{decay}< (>)$ 0, one has $m^c <(>) m^b$ and $\Delta \langle p_{\rm T} \rangle^{decay}< (>)$ 0. This feature is model independent and simply determined by the fundamental kinematics of the particles. 

\begin{table}
\renewcommand\arraystretch{1.3}
\captionsetup{justification=raggedright}
\caption{Average values of $p_{\rm T}$ and $|\eta|$ for the detected particles and their mothers in the PYTHIA event.}
\begin{tabular}{c | c | c | c | c}
& \multicolumn{2}{c|}{$\rho^0\rightarrow\pi^+\pi^-$} & \multicolumn{2}{c}{String frag. } \\
\hline
Type & \makecell[c]{unpaired \\ (case $b$, $c$)} & \makecell[c]{paired \\ (case $a$)} & \makecell[c]{unpaired \\ (case $B$, $C$)} & \makecell[c]{paired \\ (case $A$)} \\
\hline
Mother $p_{\rm T}$ & 0.75 & 0.97 & 0.94 & 1.41 \\
\hline
Mother $|\eta|$ & 1.17 & 0.53 & 2.15 & 2.12 \\
\hline
Daughter $p_{\rm T}$ & 0.59 & 0.64 & 0.68 & 0.74 \\
\hline
Daughter $|\eta|$ & 0.41 & 0.39 & 0.41 & 0.40 \\
\hline
Daughter $|\Delta\eta|$ & 1.27 & 0.48 & 1.03 & 0.69 \\
\end{tabular}
\label{tab:tab1}
\end{table}

\begin{figure*}
\centering
\includegraphics[width=\linewidth]{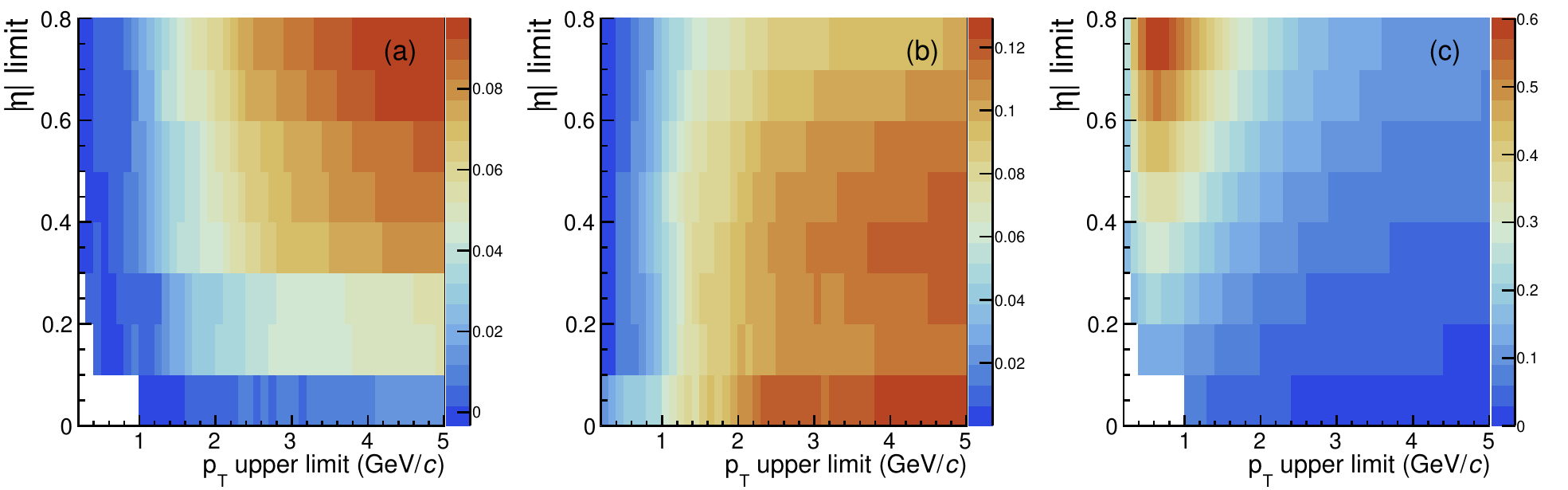}
\captionsetup{justification=raggedright}
\captionof{figure}{The slope values of $A_{\rm ch}-\Delta \langle p_{\rm T}  \rangle$ change with the $p_{\rm T}$ and $|\eta|$ coverages in the PYTHIA p-Pb collisions. (a) Tuning for both $A_{\rm ch}$ and $\langle p_{\rm T} \rangle$; (b) $p_{\rm T} > 0.2$ GeV/$c$ and $|\eta| < 0.8$ are fixed for $A_{\rm ch}$ and only tuning for $\langle p_{\rm T} \rangle$; (c) $p_{\rm T} > 0.2$ GeV/$c$ and $|\eta| < 0.8$ are fixed for $\langle p_{\rm T} \rangle$ and only tuning for $A_{\rm ch}$.}
\label{fig:slope_cuts}
\end{figure*}

In the string fragmentation scenario, the parton must carry the opposite charge with its partner for each breakup. As a result, all formed hadrons in an event , excluding the neutral ones, generally have different charges with their local neighbors, which is exactly the manifestation of the LCC as observed in Ref.~\cite{ALICE:2016}. The mechanism of Eq. (\ref{eq:5}), therefore, can be naturally extended to the primordial production as long as one treats each breakup of the string as a resonance. Figure~\ref{fig:pt_ach_str} (b) presents the local $A_{\rm ch}^{\it i}-\Delta \langle p_{\rm T}  \rangle^{\it i}$ correlations calculated string by string. Note that $B$ and $C$ only cover half of the $A_{\rm ch}$ since $A_{\rm ch}^B$ and $A_{\rm ch}^C$ are always negative and positive, respectively. An identical linear dependence is found between the string-wise $A_{\rm ch}^{\it i}-\Delta \langle p_{\rm T}  \rangle^{\it i}$ and the event-wise $A_{\rm ch}-\Delta \langle p_{\rm T}  \rangle$, suggesting the correlation is both global and local. For the same reason, the unpaired primordial hadrons have smaller $p_{\rm T}$ than the paired ones as listed in Table~\ref{tab:tab1}. Consequently, as more $h^-$ ($h^+$) are detected, the more negative (positive) $A_{\rm ch}$ is, and the lower $\langle p_{\rm T} \rangle^-$ ($\langle p_{\rm T} \rangle^+$) is. For comparison purposes, we randomly shuffle the charges of particles on the same string, which eliminates the particle wise (local) $q-p_{\rm T}$ correlation while still preserve in the string wise/event wise (global) conservation. As expected, the linear dependence vanishes, as shown in the hollow markers of Fig.~\ref{fig:pt_ach_str} (b). 

Table~\ref{tab:tab1} summarizes the mean $p_{\rm T}$, $|\eta|$ for both mothers~\footnote{Here we define the mother in the string fragmentation as the average of two partons at endpoints.} and daughters, as well as the rapidity separation $|\Delta\eta|$ between two daughters. It is found that the resonances or strings with lower $p_{\rm T}$ and/or larger $|\eta|$ tend to create unpaired particles with larger $|\Delta\eta|$, leading to the non-zero $A_{\rm ch}$. This picture agrees with the mechanisms proposed in Ref.~\cite{Bzdak:2013}. Besides $p_{\rm T}$, we also see the linear dependence between $\Delta|\eta|$ and $A_{\rm ch}$, i.e.,
\begin{equation}
\Delta |\eta| \equiv \langle |\eta^{-}| \rangle - \langle |\eta^{+}| \rangle \propto A_{\rm ch},
\end{equation}
with a negative slope on the order of $10^{-2}$, which can be experimentally measured to further examine the LCC.

To sum up, both the resonance decay and the string fragmentation are proved to follow the same pattern and together give rise to the event wise $A_{\rm ch}-\Delta \langle p_{\rm T} \rangle$ correlation. Such a relationship is the clear manifestation of the LCC, whose strength depends on the kinematic property of the string/resonance and the size of the detector acceptance.

%%%%%%%%%%%%%%%%%%%%%%%%%%%%%%%%%%%%%%%%%%%%%%%%%%%%%%%%%
\section{The impact on the search for the CMW} \label{sec:sec3}

As demonstrated, when selecting events with a specific $A_{\rm ch}$ value, in practice, one preferentially applies nonuniform $p_{\rm T}$ and $\eta$ cuts on the charged particles, resulting in the $A_{\rm ch}-\Delta p_{\rm T}$ correlation. This universal background cannot be fully subtracted and needs to be evaluated before extracting the CMW signal. Over the same $A_{\rm ch}$ range, the normalized slope of $\Delta \langle p_{\rm T} \rangle$ ($\sim$0.1) is smaller than that of $\Delta \langle v_2 \rangle$ ($\sim$0.2 - 0.3)~\cite{CMS:2019}. Considering that the $p_{\rm T}$ value is usually larger than $v_2$ by an order of magnitude, the impact of $A_{\rm ch}-\Delta p_{\rm T}$ on $A_{\rm ch}-\Delta v_2$ cannot explain the $v_2$ slope alone. Indeed, a more straightforward behavior of the LCC can be observed in the differential three-particle correlation~\cite{Voloshin:2014, ALICE:2016}.

In the experiment, one may still want to properly select the $p_{\rm T}$ range to reduce the $A_{\rm ch}-\Delta p_{\rm T}$ correlation. Generally, the narrower the $p_{\rm T}$ range is, the less this effect is included. On the other hand, a wider $p_{\rm T}$ range enhances particle yields. Hence, we suggest that the integrated $v_2$ in different $A_{\rm ch}$ bins can be scaled by its $\langle p_{\rm T} \rangle$ no matter which $p_{\rm T}$ range is chosen. Moreover, the $\Delta v_2$ can be experimentally obtained in two ways: find the $p_{\rm T}$-integrated $v_{n}$ in a given $p_{\rm T}$ range for $h^-$ and $h^+$, and then take the difference, or start with the $v_{n}$ difference between $h^-$ and $h^+$ as a function of $p_{\rm T}$, and then fit the difference in a given range with a constant to get the average. Ideally, these two methods should be consistent, however, in the presence of the already known LCC background, the second way should be more appropriate since it minimizes the $\Delta v_2$ induced by $\Delta \langle p_{\rm T} \rangle$ in each $p_{\rm T}$ bins and is sensitive to any fluctuation of the $v_2(p_{\rm T})$.

The slope caused by the LCC and by the CMW may behave differently in the varied kinematic windows. Figure~\ref{fig:slope_cuts} presents how the slope of $A_{\rm ch}-\Delta \langle p_{\rm T}  \rangle$ changes with the $p_{\rm T}$ and $|\eta|$ coverages in the PYTHIA model. It can be seen in three panels, respectively, that: (a) when narrowing down the upper limits of $p_{\rm T}$ and $|\eta|$ for both $A_{\rm ch}$ and $\langle p_{\rm T} \rangle$, the slope is gradually decreased; (b) when $p_{\rm T} > 0.2$ GeV/$c$ and $|\eta| < 0.8$ are fixed for $A_{\rm ch}$, the slope is slightly increased as the $|\eta|$ range for $\langle p_{\rm T} \rangle$ decreases, and decreased as the $p_{\rm T}$ range for $\langle p_{\rm T} \rangle$ decreases; and (c) when $p_{\rm T} > 0.2$ GeV/$c$ and $|\eta| < 0.8$ are fixed for $\langle p_{\rm T} \rangle$, the slope is dramatically increased by a factor of 6 as the $p_{\rm T}$ range for $A_{\rm ch}$ decreases to 1 GeV/$c$, and decreased as the $|\eta|$ range for $A_{\rm ch}$ decreases. If the measured $A_{\rm ch}-\Delta v_2$ is merely due to the LCC, one would expect the synchronous change between $A_{\rm ch}-\Delta v_2$ and $A_{\rm ch}-\Delta \langle p_{\rm T} \rangle$. By comparison, we directly calculate the $A_{\rm ch}-\Delta v_2$ relation with the AMPT model initially imported the quadrupole configuration~\cite{Ma:2014, Shen:2019} but lacking of the LCC dynamic and find that the slopes only vary moderately in these kinematic windows. It is therefore worthwhile to experimentally compare these slopes to disentangle the LCC- and the CMW-induced $A_{\rm ch}-v_2$ relation. A preliminary STAR result~\cite{Shou:2019} partially examining the slopes in varied $p_{\rm T}$ ranges is consistent with our simulation in Fig.~\ref{fig:slope_cuts} (b) and further measurements would be more helpful .

\begin{figure}
\centering
\includegraphics[width=0.9\linewidth]{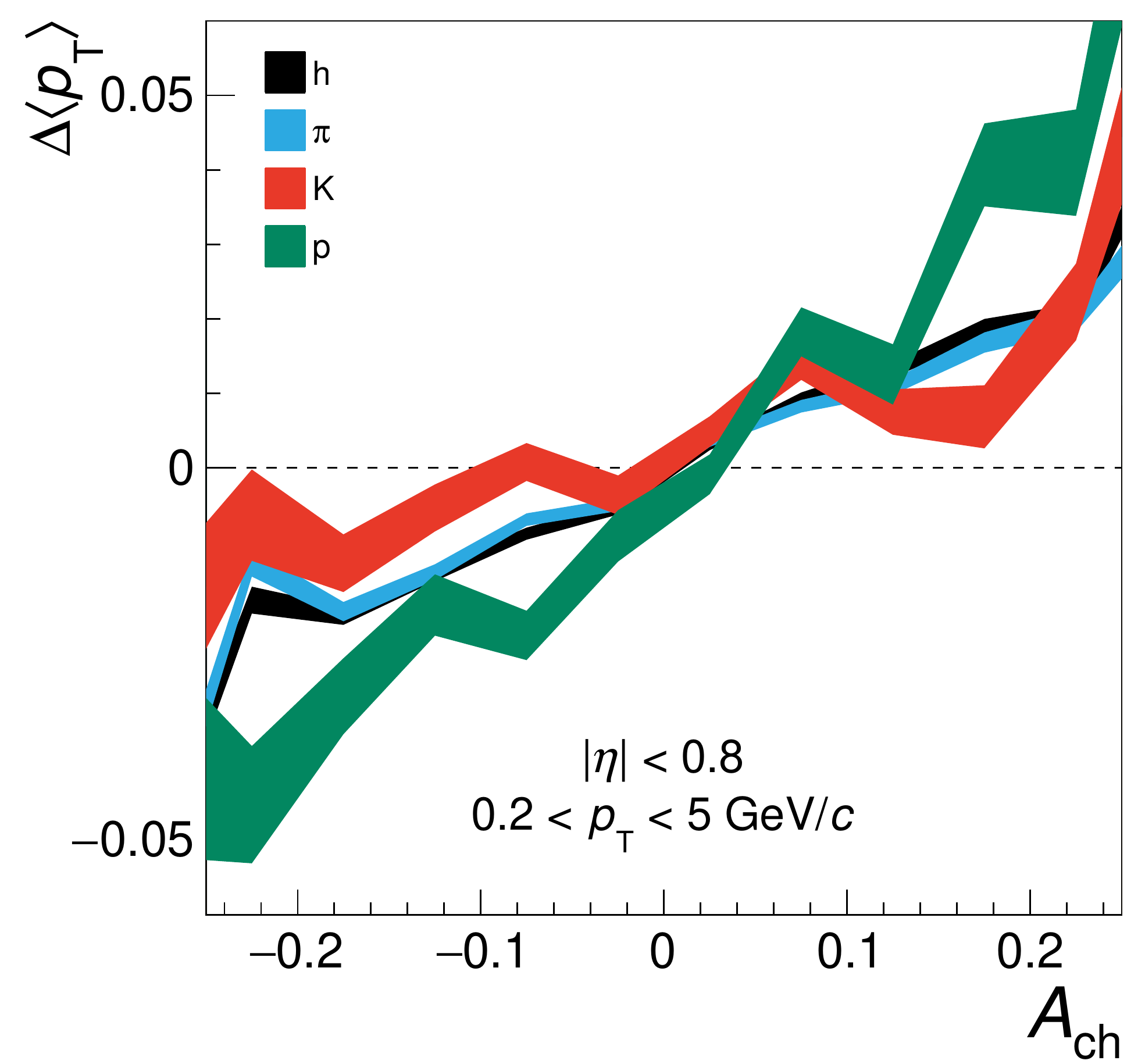}
\captionsetup{justification=raggedright}
\captionof{figure}{Dependence of $\Delta \langle p_{\rm T} \rangle$ on $A_{\rm ch}$ for $h$, $\pi$, $K$ and $p$ in the PYTHIA simulation.}
\label{fig:pt_ach_pid}
\end{figure}

Another interesting measurement would be with identified hadrons. The CMW is originally theorized to affect only light quarks~\cite{Burnier:2011} and its flavor dependence remains unclear. The slope for kaons is suggested to be negative in Ref.~\cite{Hatta:2016} because the isospin chemical potentials between $K$ and $\pi$ are opposite. It has been tentatively negated, however, by the STAR preliminary data~\cite{Shou:2019}. In the perspective of the LCC, all charged hadrons regardless of species follow the universal $A_{\rm ch}-\Delta \langle p_{\rm T}  \rangle$ correlation. Thus, the slopes of pions, kaons and protons are expected to be similar and positive ($\sim$0.1), as shown in Fig.~\ref{fig:pt_ach_pid}.

One should be aware that the anisotropic flow in PYTHIA and HIJING are very small. This makes it infeasible to directly examine the $A_{\rm ch}-v_2$ correlation without additional modifications. In contrast, AMPT model, which succeeds in describing the collectivity, lacks the necessary LCC environment as shown in Fig.~\ref{fig:pt_ach} (d). A recent simulation~\cite{Zhao:2020} claims that the HIJING model, when properly scaled, is able to reproduce the behavior of the $\gamma$ correlator in the CME study. For this reason, we speculate that the string fragmentation models, when scaled by the correct flow parameters, can also quantitatively describe the experimental measurement of the CMW, which is worth a try in future studies.

%%%%%%%%%%%%%%%%%%%%%%%%%%%%%%%%%%%%%%%%%%%%%%%%%%%%%%%%%
\section{Summary} \label{sec:sum}

The CMW has been experimentally sought in heavy-ion collisions through the $A_{\rm ch}$ dependence of $v_2$. Since $v_2$ linearly depends on $p_{\rm T}$, the $A_{\rm ch}$ dependent $\langle p_{\rm T} \rangle$ could naturally play a role as a background. With the string fragmentation models, including PYTHIA, DPMJET and HIJING, we quantitatively reproduce the $A_{\rm ch}-\Delta \langle p_{\rm T} \rangle$ correlation observed in the data. Such an event-wise correlation can be traced back to the local level. When dissecting the event into different production cases (strings and resonances), it is found that the event $A_{\rm ch}$ is mainly determined by the numbers of each case. The key mechanism leading to the $q-p_{\rm T}$ relation is exactly what the LCC implies; namely, when particles are produced in charge-conserving pairs, the unpaired hadron whose partner is excluded by the limited acceptance usually carries smaller $p_{\rm T}$ compared to the paired hadrons. As more unpaired $h^-$ ($h^+$) are detected, the more negative (positive) $A_{\rm ch}$ is, and the lower $\langle p_{\rm T} \rangle^-$ ($\langle p_{\rm T} \rangle^+$) is. Both string fragmentation and the resonance decay are proven to follow the same scenario and to generate the similar positive slopes. We argue that when selecting events with a specific $A_{\rm ch}$, in practice, one preferentially applies nonuniform $p_{\rm T}$ and $\eta$ cuts on the charged particles and such a LCC background is too ubiquitous to be fully eliminated. We also propose that measuring the slope of $A_{\rm ch}-\Delta  p_{\rm T}$ $(\Delta|\eta|)$ at varied kinematic windows and with identified hadrons may shed more light on disentangling the difference between the LCC-induced and the CMW-induced $A_{\rm ch}-v_2$ dependence.

%%%%%%%%%%%%%%%%%%%%%%%%%%%%%%%%%%%%%%%%%%%%%%%%%%%%%%%%%
\section*{Acknowledgments}

We are grateful to A. H. Tang, S. A. Voloshin, G. Wang, F. Wang and H.-J. Xu for the enlightening discussions and suggestions. We also thank W.-B. He, G.-L. Ma, S. Zhang and C. Zhong for their assistance. This work is supported by the Strategic Priority Research Program of Chinese Academy of Sciences (No. XDB34030000), the National Natural Science Foundation of China (Nos. 11890710, 11890714, 11975078, 11421505, 11605070) and the National Key Research and Development Program of China (Nos. 2016YFE0100900, 2018YFGH000173). QS is sponsored by the Shanghai Rising-Star Program (20QA1401500).

%%%%%%%%%%%%%%%%%%%%%%%%%%%%%%%%%%%%%%%%%%%%%%%%%%%%%%%%%
{}

\end{document}